# A General Approach for Using Deep Neural Network for Digital Watermarking

Yurui Ming, Weiping Ding, *Senior Member, IEEE*, Zehong Cao, and Chin-Teng Lin, *Fellow, IEEE*

*Abstract*—Technologies of the Internet of Things (IoT) facilitate digital contents such as images being acquired in a massive way. However, consideration from the privacy or legislation perspective still demands the need for intellectual content protection. In this paper, we propose a general deep neural network (DNN) based watermarking method to fulfill this goal. Instead of training a neural network for protecting a specific image, we train on an image set and use the trained model to protect a distinct test image set in a bulk manner. Respective evaluations both from the subjective and objective aspects confirm the supremacy and practicability of our proposed method. To demonstrate the robustness of this general neural watermarking mechanism, commonly used manipulations are applied to the watermarked image to examine the corresponding extracted watermark, which still retains sufficient recognizable traits. To the best of our knowledge, we are the first to propose a general way to perform watermarking using DNN. Considering its performance and economy, it is concluded that subsequent studies that generalize our work on utilizing DNN for intellectual content protection is a promising research trend.

*Index Terms*— Deep Neural Network (DNN), Digital Content Protection; Digital Watermarking; Privacy; Internet of Things (IoT)

## I. Introduction

Technologies related to the Internet of Things (IoT) undoubtedly accelerate the speed and volume in digital content acquisition [1]. An obviously example is the ubiquitous cameras in the variety of circumstances, such traffic monitoring [2], assembly line inspection [3], environmental hazard detection [4], to name a few, can capture images in a massive volume. However, the reduction of cost on the digital content acquisition in no way compromises the importance of the content protection. For instance, in the pursuit of traffic violation, the police should present scene images which are intact or authentic in supporting the case [5]. Therefore, research on methods in this regard is of persistent interests for researchers.

Among the technologies which facilitate embedding information into digital content for authentication or protection, digital watermarking is a widely and actively used method [6-8]. In this paper, we focus our research on the image case with invisible digital watermarking. Currently, the watermarking techniques are mainly divided into two categories. The approaches in the first category are carried out in the spatial domain, such as manipulating the least significant bits [9], and patch-based methods [10]. The advantage is that these methods are simple to implement; however, they are not resistant to operations applied to the watermarked image, such as filtering, transform and re-quantization, etc. The second category of methods work on the frequency domain or transformed domain. For example, to embed the digital secrecy into the intermediate frequency components of the image after transformed from the spatial domain via discrete cosine transform (DCT) or discrete wavelets transform (DWT) [11-14]. Although these methods are robust to manipulations of the watermarked image, they are complicated in implementation.

There are several criteria that must be satisfied for a method to be considered for digital watermarking. These criteria are mainly from the perceptive and robust perspectives, in addition to other aspects such as non-removal and unambiguity. Perceptive aspect states that the embedded information must not be perceived in an obvious and subjective way, even under intended manipulation. This is critical especially for invisible watermarks. The robust criterion requires that the watermarked image must resist common filtering operations such as blurring and enhancing to retain the secretory information [15], no matter these operations occur in spatial domain or frequency domain. These criteria are also considered when we evaluate our proposed method.

Yurui Ming and Chin-Teng Lin are with the School of Computer Science, Centre for Artificial Intelligence, University of Technology Sydney, NSW 2007 Australia (e-mail: yrming@gmail.com, Chin-Teng.Lin@uts.edu.au).

Weiping Ding is with the School of Computer Science and Technology, Nantong University, Nantong 226019, China (e-mail: ding.wp@ntu.edu.cn).

Zehong Cao is with the Department of Discipline of ICT, School of Technology, Environments and Design, University of Tasmania, Hobart, TAS 7001, Australia (e-mail: zhcaonctu@gmail.com).

Essentially, digital watermarking that requires an invertible complicated method to embed information into the target image are generally non-linear, regardless the domains. Hence, neural network especially the deep neural network (DNN) that exhibits high linearity, can be a candidate approach. There already exists utilization of neural network for digital watermarking [16-19]; however, these approaches mainly address how to tailor a neural network for protecting one particular digital image. Considering the overhead for training neural network especially DNN, it is hardly to afford the cost for training individual neural network for protecting each image in a massive way, and the conventional treatments present great challenge and impractical for real applications.

However, DNN which succeeds in various applications shed the potentiality for digital watermarking [20]. The complexity of deep network structure indicates the possibility of blending the secret information and target images in a more intangible but appropriate way; and the general monotonic activation function indicates the invertible process to retrieve hidden information from the watermarked image [21].

In this paper, by virtue of the DNN, we design a DNN architecture to fulfil digital watermarking in a general and economic manner. The paper is structured with our contributions. In section II, we describe the motivation and design the corresponding DNN model. The overall treatment is to train the model on an image set first, then after the network learns how to embed the watermark into original images and retrieve the watermark from the watermarked images, the model is tested against distinct images to verify the generalizing capabilities. In section III, by instantiating the network according to a certain configuration, the proposed method is evaluated on a public image dataset and intriguing results are illustrated. In section IV, the method is systematically assessed by referring to the criteria of watermarking to show the competence of our approach. To the best of our knowledge, we are the first to utilizing DNN for watermarking in a general way and confident of further generalization.

## II. MOTIVATION AND ARCHITECTURE

As aforementioned, for digital watermarking, it requires a method or series of operations to embed the watermark into a targeted image and retrieve the watermark when needed as in Fig. 1. Watermarking can be regarded as a complicated computation process with extra properties, such as inversible. From the mathematical perspective, denote the image as $I$, the watermark as $w$, the aim of watermarking is to seek an operation $G$ that satisfies:

$$\bar{I} = G(I, w) \wedge \|\bar{I} - I\| < \varepsilon. \tag{1}$$

$\bar{I}$ is the watermarked image. The inverse property of $G$ means that there exists $G^{-1}$ satisfying $(I, w) \approx G^{-1}(\bar{I})$. Usually, upon satisfying $\|\bar{I} - I\| < \varepsilon$, the de-watermarking process tends to put emphasis on the retrieved watermark. Hence, the conventional watermarking can be alternatively written as:

$$w = G_I^{-1} G_I(w) \tag{2}$$

i.e., each $G$ is parameterized by an individual image $I$. This explains that conventional watermarking is quite specific.

Recent years witness the great achievement of DNN in various applications. It also revolutionizes some conventional fields with supervising accomplishments, such as GAN (generative adversarial network) [22], DQN (deep Q-network) [23], etc. The computational capability of the DNN, either as a mapping from the uniform random distribution to a specific random distribution, or as a powerful function estimator, demonstrate the superior competence in a variety of tasks. In these applications, no explicit rules delivered to DNN to guide its behavior, and DNN learns from samples or experiences, and generalize to new situation. Therefore, considering the powerfulness of DNN, it is appealing to consider its potentiality for digital watermarking.

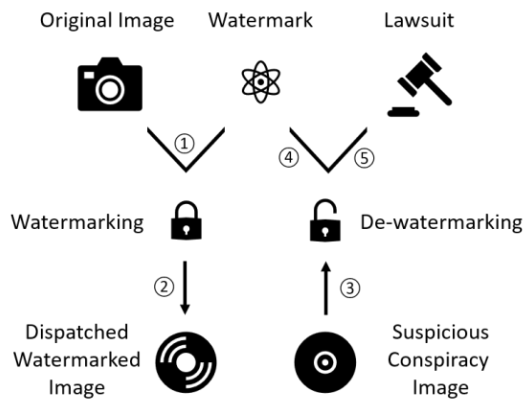

Fig. 1. Paradigm of digital watermarking for images. Note ⑤ is dependent on ④, the assessment of the retrieved watermark.

A retrospection of the general application of DNN reveals that the paradigm is rather stereotypical, i.e., training the designed network in a training set and test it on a separate test set. Hence, a straightforward migration of DNN to digital watermarking is similar, i.e., training the designed network on substantial images to enable the network to learn the way for embedding and retrieving of the watermark image; then test the learned capability of network on other distinct images. Extra aspect is also needed to be manifested, such as the requirement of training of DNN in an end-to-end way.

Furthermore, the above idea of utilizing DNN for watermarking might be potentially compliant with the theoretic analysis. Previously we show that conventional watermarking can be represented as in (2). If $G$ is implemented by DNN, i.e., parameterized by weights $\theta$, and we can tactically train $G$ to have optimal $\theta$ that is common to all images, thus $G$ can be parameterized by $\theta$ in a latent but general manner:

$$w = G_\theta^{-1} G_\theta(w) \qquad (3)$$

Comparison of (2) and (3) indicates this reparameterization frees $G$ from dependent on a specific image but features common to an image collection. It releases the potentiality that $G_\theta$ can be utilized to watermarking other images.

Based on the above description, the overall architecture of the designed neural network is in Fig. 2. It consists of several modules or subnetworks that work together to fulfil the final goal. We first utilize transpose convolution [24] to convert the original image and watermark image into a higher dimensional space to blend them together. The module for up-lifting of the dimension is named up-sampler, and the mix-up operation is done the module named blender, both are neural networks. After embedding the copywrite information, a component named down-sampler is made use of to have the blended image to be with the same dimension of the original image, as well as restore some features after blending. In order to assess that a digital artwork is protected via watermarking, an extracting sub-network is designed to retrieve the embedded information.

The reason for designing the up-sampler module is as follows. As mentioned above, the watermarking in transformed domain is effective; however, it is not intuitive to design a transformed domain with distinctive properties for neural network. However, we postulate that a higher dimensional space (or latent space) resulted from network operations might resemble some similarities, for example, it has a higher freedom to blend the pixels from the original image and watermark image. This freedom might be beneficial for network since it can choose the most appropriate way by learning to increase the quality of embedding and resist manipulations to the watermarked image.

The functionality and performance of the model are measured by comparing the original image with watermarked image, the original watermark with the extracted watermark simultaneously according to (4), which is in form of the mean square error:

$$L(\theta) = \sum_{i,j} (\bar{A}_{i,j}(\theta) - A_{i,j})^2 + \sum_{i,j} (\bar{W}_{i,j}(\theta) - W_{i,j})^2 \qquad (4)$$

$A$, $W$, $\bar{A}$ and $\bar{W}$ denote the original image, original watermark, watermarked image and extracted watermark respectively. Notably, in (1), $\theta$ represents the overall parameters, although the parameters for $\bar{A}$ is a subset of the parameters on which $\bar{W}$ depends. Furthermore, for the blender network, the weights $w_s$ and $w_e$ in Fig. 1 are also trainable variables by design, which automatically take account of the importance of watermark image pixels when blending.

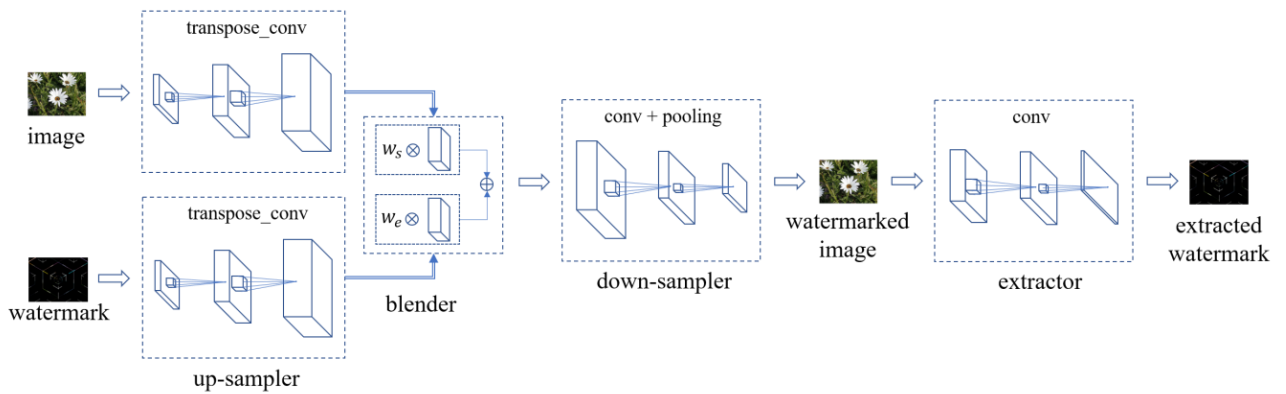

Fig. 2. Network architecture for general digital watermarking (The watermark image can be referred to Fig. 3(C) for a better illustration).

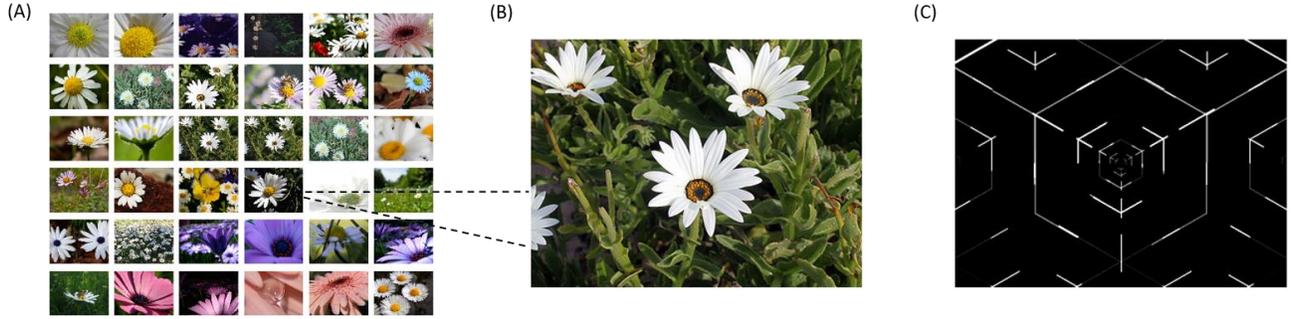

Fig. 3. Images for training and the corresponding watermark image (note the watermark image is rescaled with the maximal pixel value from 63 to 255 for better illustration).

## III. EXPERIMENTS

For convenience and avoidance of copywrite infringement, we utilize a dataset from Kaggle [25]. The dataset is a collection of images for flowers recognition. The preference of this dataset also lies in other considerations. For example, the images in the dataset are about 320x240 pixels, a reasonable resolution for carrying out the experiment. In addition, the images are categorized into five categories; we can use the first four categories of images for training, and the last category for testing. The distinctiveness of training images and test images is a stronger evidence to show the practicability of the proposed network architectures on success.

These images and the watermark are shown in Fig. 3. Fig. 3(A) is a snapshot of the training images. The canonical dimensions for images to the network are set to 320x240, so the images are selected and manipulated to match the dimension constraint. After rectifying the images, there are 1703 images for training and 427 images for test. A watermark image collected from Internet is shown in Fig. 3(C). We are in courtesy of the original provider and with no intention to infringe copyright besides the sole research purpose of this paper. The watermark is chosen to be the same size of training image to ease network operations, hence the result could be further improved for carefully tweaked watermark in smaller size.

The instantiation of the architecture is via TensorFlow [26]. It uses the de facto modules provided by the library and no more customized operation for easily replication of this work to benefit subsequent research. The configuration of the network architecture is shown in TABLE I.

With the above configuration, we train the neural network for 10,000 iterations with batch size 8 and a learning rate of 0.001. To stabilize the training process, the learning rate is decayed for every 400 iterations by a factor of 0.92. To avoid overfitting during the training process, we take a batch number of 8 from test image set to monitor the training process. In theory, the image statistical distributions of training set and test set are different from each other, it is more objective to assess the training process. The feasibility of our proposed method can be preliminarily demonstrated by the validation process as in Fig. 4.

## IV. EVALUATIONS

### A. Watermarked Images

To systematically assess the quality of watermarked images, we evaluate them from two perspectives. The first is from subjective aspect. We select 6 images from the test set at random and present them to 6 university students. For images with simple and monotonous textures, it is reported tiny perceivable difference between the original image and watermarked image. For images with intermediate complex textures or scenarios, the perceivable difference is neglectable and only being noticed upon presence of the watermark image as reminder. For images with highly complex textures or scenarios, there is no subjective difference even with presence of the watermark image as reminder. We show some cases in Fig. 5 and Fig. 6 respectively. As the first attempt in performing watermarking in a general way, the subjective results illustrate the potentiality of considering more tricks in neural network to improve the perfection.

TABLE I
NETWORK CONFIGURATION

| Name | Operation | #Features | Filter Size | Stride | Comments |
|---|---|---|---|---|---|
| Up-sampler | Conv2DTranspose | 16 | 5x5 | 2 | No bias |
| | Conv2DTranspose | 16 | 5x5 | 2 | No bias |
| | Conv2DTranspose | 16 | 5x5 | 2 | No bias |
| Down-sampler | Conv2D | 12 | 5x5 | 1 | No bias |
| | AvgPool2D | - | 2x2 | 1 | |
| | Conv2D | 6 | 5x5 | 1 | No bias |
| | AvgPool2D | - | 2x2 | 1 | |
| | Conv2D | 3 | 5x5 | 1 | No bias |
| | AvgPool2D | - | 2x2 | 1 | |
| Extractor | Conv2D | 12 | 5x5 | 1 | No bias |
| | Conv2D | 6 | 5x5 | 1 | No bias |
| | Conv2D | 3 | 5x5 | 1 | No bias |

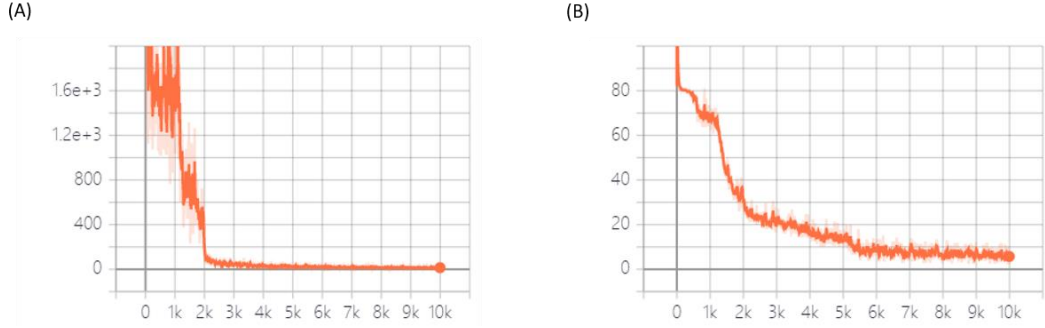

Fig. 4. Validation loss during the training process (A) image loss; (B) watermark loss.

To objectively assess our method, we adopt the PSNR (Peak Signal-to-Noise Ratio) defined in [27] to measure the confidence objectively:

$$PSNR = -10 \cdot \log_{10} \frac{\sum_{k=1}^{3}\sum_{m=1}^{M}\sum_{n=1}^{N}(\bar{A}_k(m,n)-A_k(m,n))^2}{3*255^2*M*N} \tag{5}$$

In [27], it reported that PSNR larger than 38 dB is regarded as the good quality of the watermarked image (In [28], the threshold of PSNR is recommended 30 dB). The PSNR of the 6 watermarked images chosen at random are in TABLE II, with an average of 38.6. However, PSNR might not in fully compliance with the subjective perception. Fig. 6 shows the cases of watermarked image with the lest PSNR, i.e., 32.1, the intrinsic texture of the image still prohibits an easy discrimination. Overall, the objective assessment confirms the promising adoption of this generic method.

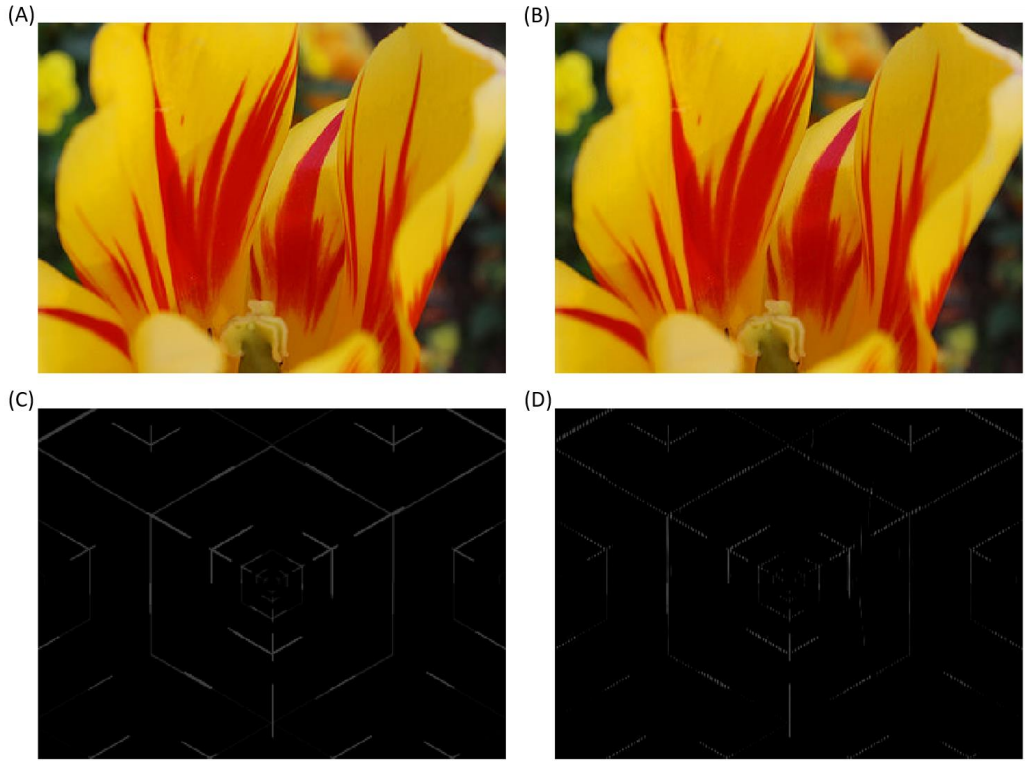

Fig. 5. Case for an image with simple and monotonous textures. The perceptive differences can be spotted only from special observation angles, but they are overall still quite tiny. (A) Original image; (B) Watermarked image; (C) Original watermark; (D) Extracted watermark.

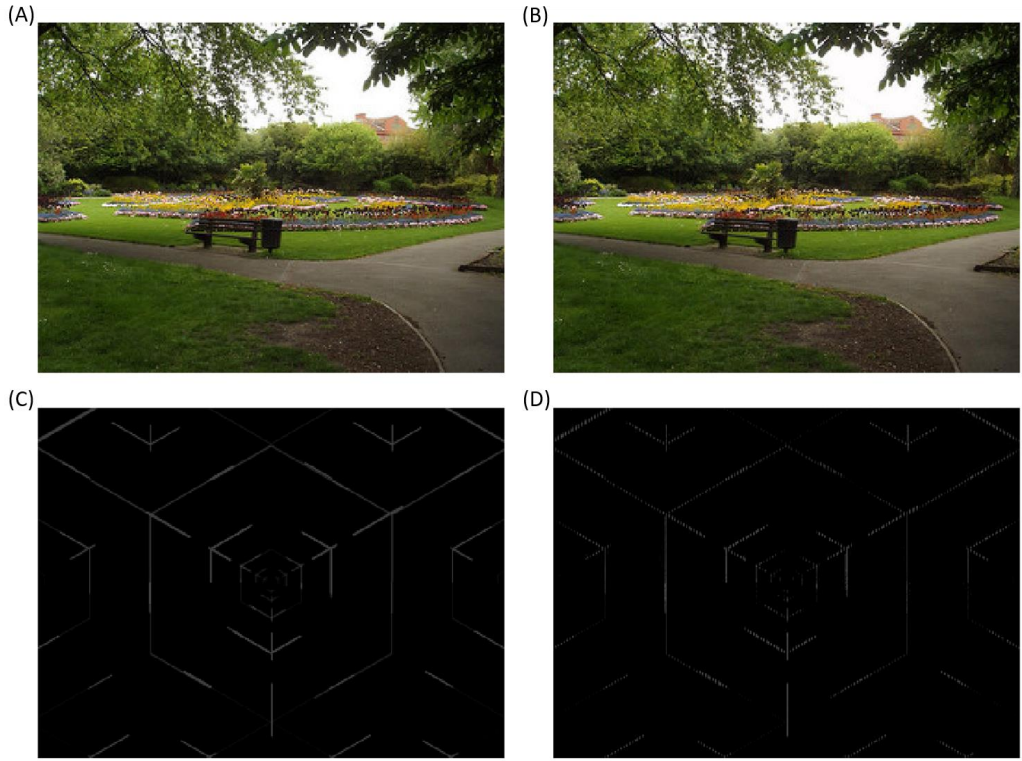

Fig. 6. Case for an image with highly complicated texture and scenario. The perceptive difference is invisible to observers even with the presence of watermark image as a reminder. (A) Original image; (B) Watermarked image; (C) Original watermark; (D) Extracted watermark.

TABLE II
SNR OF WATERMARKED IMAGES

| IMG ID | 123 | 99 | 174 | 333 | 396 | 294 | Average | Baseline [27] |
|---|---|---|---|---|---|---|---|---|
| PSNR* | 41.8 | 39.1 | 38.1 | 36.2 | 32.1 | 44.4 | 38.6 | 38 |

*units: dB

## B. Extracted Watermark

To assess the robustness of the proposed watermarking mechanism, several modifications are applied to the watermarked images to examine the extracted watermark. Same as the evaluation of watermarked image, the assessment of watermark is categorized into subjective and objective way. The modifications considered here includes clipping, low-pass filtering, high-pass filtering, noise degrading.

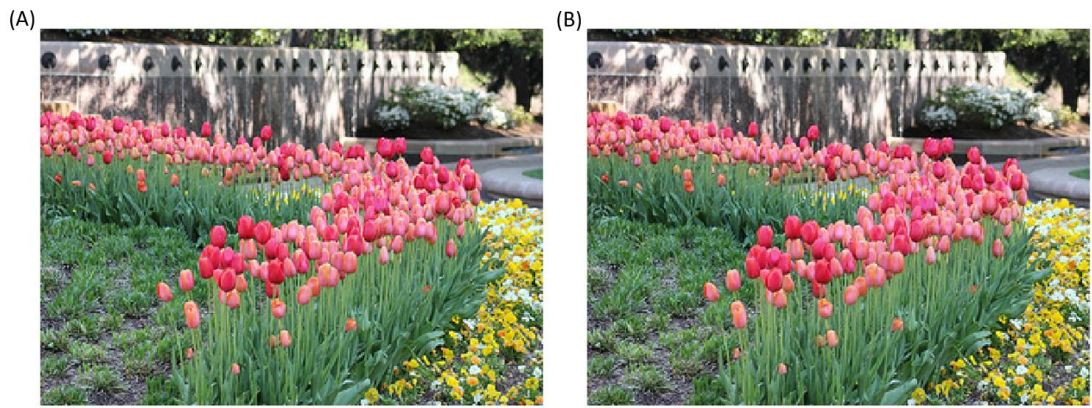

Fig. 7. Image with intrinsic complex context reporting low PSNR of 32.1 still prohibits easy discrimination between the original image and watermarked image. (A) Original image; (B) Watermarked image.

| (A) | | | (B) | | |
|---|---|---|---|---|---|
| 0.0625 | 0.125 | 0.0625 | −0.125 | −0.25 | −0.125 |
| 0.125 | 0.25 | 0.125 | −0.25 | 2.5 | −0.25 |
| 0.0625 | 0.125 | 0.0625 | −0.125 | −0.25 | −0.125 |

Fig. 8. (A) Gaussian filter; (B) Laplacian filter.

For clipping, part of the watermarked image is chopped from the image and compensated with zero to keep the original size for watermark extraction. For low-pass filtering, we apply a Gaussian filter in the spatial domain, while high-pass filtering is carried out by applying a Laplacian filter, as in Fig. 8. For noise degrading, a random Gaussian noise is merged into the watermarked image to inspect the affect to extracted watermark.

For subjective assessment, we find that besides low-pass filtering, all other operations can retain the watermark in sufficient quality from the perceptive perspective. We only illustrate the case of high-pass filtering in Fig. 9, whilst the case of low-pass filtering in Fig. 10. We will defer the discussion of vulnerability of this method to the blurring modification, however, the root cause is still planned as future work.

To objectively assess the extracted watermark after various modifications to the watermarked image, we adopt the measurement in [28], i.e., the normalized correlation (NC) in (6):

$$NC = \frac{W \cdot \overline{W}}{\sqrt{\|W\|^2}\sqrt{\|\overline{W}\|^2}} \quad (6)$$

where

$$\|W\|^2 = \sum_{k=1}^{3}\sum_{m=1}^{M}\sum_{n=1}^{N} w(m,n)^2 \quad (7)$$

$$W \cdot \overline{W} = \sum_{k=1}^{3}\sum_{m=1}^{M}\sum_{n=1}^{N} w(m,n) * \overline{w}(m,n) \quad (8)$$

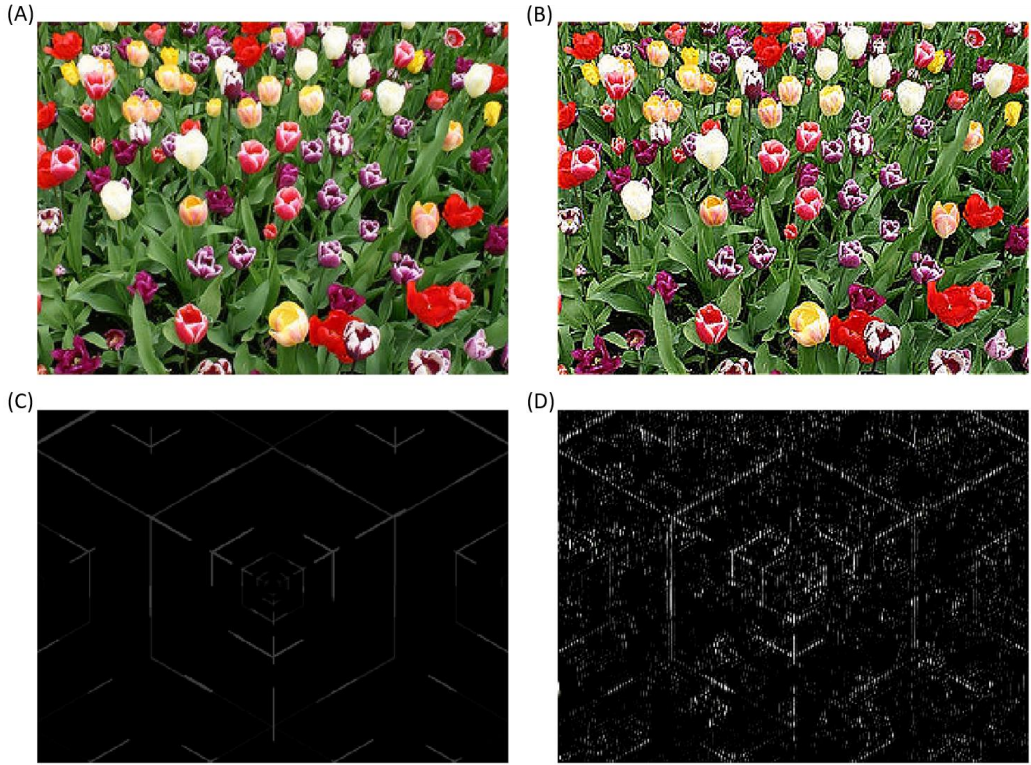

Fig. 9. Case for the high-pass filtering. Although there are changes for the extracted watermark, however, the overall contours are still rather perceivable. (A) Watermarked image; (B) (High-pass) filtered image; (C) Original watermark; (D) Extracted watermark.

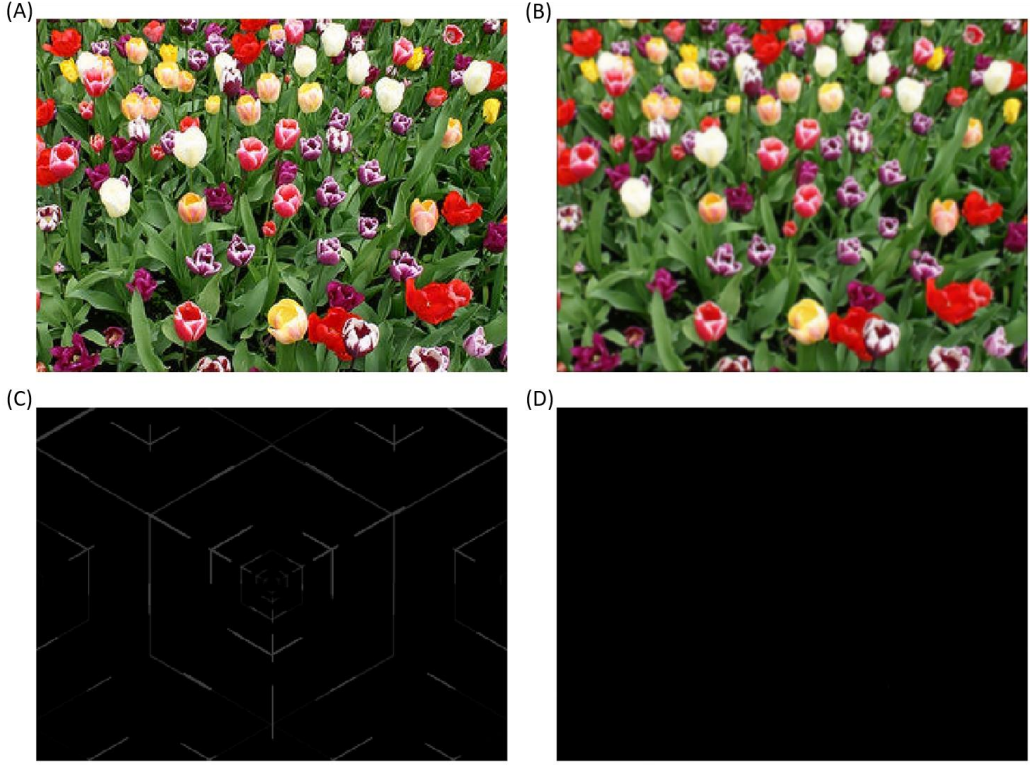

Fig. 10. Case for the low-pass filtering. The network fails in extracting the watermark. (A) Watermarked image; (B) (Low-pass) filtered image; (C) Original watermark; (D) Extracted watermark.

$W$ denotes the original watermark, and $\overline{W}$ denotes the watermark extracted from the modified watermarked image.

Notably, this measurement is implicitly assumed that the NC between the extracted watermark from the untouched watermarked image and original watermark is 100%. However, it is usually not the case. Whether the watermarked image undergoes some modification or not, the watermark tends to exhibit some degradation after extraction. Denote the NC between original watermark and the watermark extracted from the intact watermarked image as $NC_0$, we rewrite (6) as (9):

$$NC = \frac{W \cdot \overline{W}}{\sqrt{\|W\|^2}\sqrt{\|\overline{W}\|^2}} \cdot \frac{1}{NC_0} \qquad (9)$$

Based on (9), we calculate the NC as in TABLE III, where $NC_0$ equals 70.34%. However, there is no literature explicitly about the threshold of the NC. But according to our experience, 50% can be qualified especially from the perceptive perspective. TABLE III indicates the commonly-used operations, such as clipping and sharpening (enhancement), exhibit satisfying NC. However, further improvement of the NC for corresponding modifications is planned as our future work.

TABLE III
NC OF EXTRACTED WATERMARK FOR A GIVEN IMAGE

| OPS | Clipping | Low-pass Filtering | High-pass Filtering | Noise Degrading |
|---|---|---|---|---|
| NC* | 74.42 | 4.22 | 64.89 | 49.23 |

*units: percentage

## V. DISCUSSION

To the best of our knowledge, we are the first to introduce DNN to perform digital watermarking in a general way. Therefore, there exists several aspects that need further study. For example, how to architect more elegant DNN to fulfill the watermarking task, meanwhile improve the robustness for modifications against watermarked image. In this discussion, we try to address the concern we raised in the last section, i.e., the potential reason that low-pass filtering induces dramatic impact on the watermarked image. However, the preliminary results above indicate the promising achievements if subsequent research could be carried out in this regard.

In order to understand the low-pass filtering affection, it is necessary to look into the characteristic of the sub-network, aka, extractor, which is used to extract watermark from the watermarked image. It is indeed a deep convolutional network; hence we only need to investigate the traits of these weights. For convenience, we focus the weights (or kernels) of the last convolutional layer.

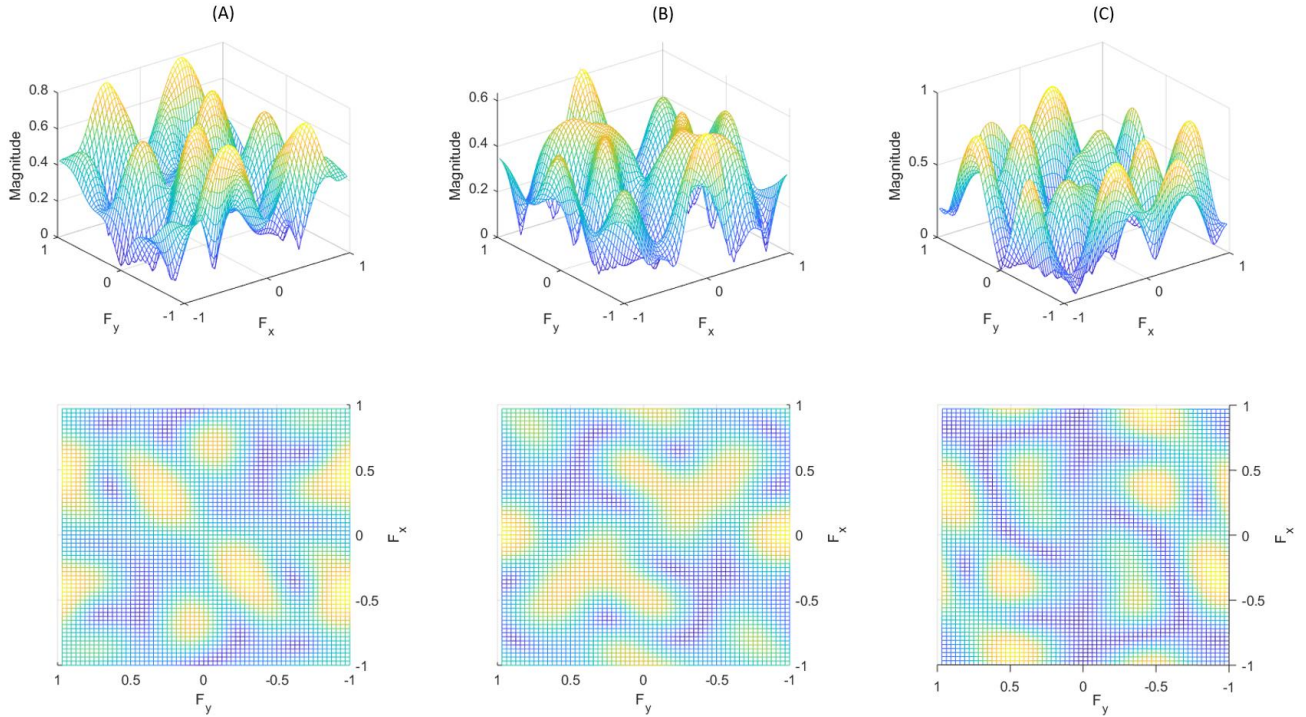

Fig. 11. Frequency responses of the clustered kernel centroids. The bottom figure of each column is the rotated version of the top figure.

However, even for the last layer, there are altogether 18 kernels. It is time consuming to investigate all these kernels one-by-one. Therefore, we perform a clustering on these kernels via K-means [29], to partition them into 3 categories and study their centroids. Fig. 11 shows the frequency responses of the centroids.

From these responses, especially the bottom figures, it is manifest that the kernels are potentially centrosymmetric. They might not be strictly high or bass pass filters, however, they are obviously not low-pass filters. Along a specific direction, most of them can be regarded as high-pass filters. They might suggest that the extractor network is by "sharpening" the watermarked image to extract the buried watermark information. But if the watermarked image undergoes a low-pass filtering, this might counter the effectiveness of extraction. The good news is deliberate blurring of image is not very common image processing operation; however, in future work we will study more elegant network architectures and tricks to improve the robustness of this general approach.

## VI. Conclusion

In this paper by considering the paradigm of utilizing DNN and the accomplishments it made, we proposed a general way in applying DNN for digital watermarking. By constructing a DNN which suitable for the problem and trained on a set of images, the experiment on test images revealed the potential in this manner. The subjective and objective assessments both demonstrate the practicability and economy of this approach. To the best of our knowledge, we are the first to carry out a general way for utilizing DNN for digital watermarking, and this preliminary achievement manifests the meaningfulness for conducting further research in this thread.

ACKNOWLEDGMENT

This work was supported in part by the Australian Research Council (ARC) under discovery grant DP180100670 and DP180100656; NSW Defense Innovation Network and NSW State Government of Australia under the grant DINPP2019 S1-03/09; Office of Naval Research Global, US under Cooperative Agreement Number ONRG-NICOP-N62909-19-1-2058.